\documentclass[conference,compsoc]{IEEEtran}
\usepackage{mathptmx} 

\usepackage{comment}

\usepackage{amsmath,amssymb,amsfonts}
\usepackage{graphicx}
\usepackage{textcomp}
\usepackage{xcolor}

\usepackage{pdfpages}
\usepackage{listings}
\usepackage[hyphens]{url}
\usepackage{fancyhdr}
\usepackage[normalem]{ulem}
\usepackage[hyphens]{url}
\usepackage[sort,nocompress]{cite}
\usepackage[final]{microtype}
\usepackage[bookmarks=true,breaklinks=true,letterpaper=true,colorlinks,linkcolor=black,citecolor=blue,urlcolor=black]{hyperref}

\usepackage{xcolor}
\usepackage{subcaption}
\usepackage{makecell}
\usepackage{enumitem}
\usepackage[]{footmisc}

\begin{document}

\title{RollingCache: Using Runtime Behavior to Defend Against\\ Cache Side Channel Attacks}

\author{
\IEEEauthorblockN{Divya Ojha}
\IEEEauthorblockA{Department of Computer Science\\
University of Rochester\\
divya.ojha.263@gmail.com}
\and
\IEEEauthorblockN{Sandhya Dwarkadas}
\IEEEauthorblockA{Department of Computer Science\\
University of Virginia\textsuperscript{\textsection}\\
sandhya@virginia.edu}
}


\maketitle
\begingroup\renewcommand\thefootnote{\textsection}
\footnotetext{Work done while at University of Rochester.}
\endgroup

\begin{abstract}

Shared caches are vulnerable to side channel attacks through contention in cache sets. Besides being a simple source of information leak, these side channels form useful gadgets for more sophisticated attacks that compromise the security of shared systems. 

The fundamental design aspect that contention attacks exploit is the deterministic nature of the set of addresses contending for a cache set. 
In this paper, we present RollingCache, a cache design that defends against contention attacks by dynamically changing the set of addresses contending for cache sets.
Unlike prior defenses, RollingCache does not rely on address encryption/decryption, data relocation, or cache partitioning. 
We use one level of indirection to implement dynamic mapping controlled by the whole-cache runtime behavior.
Our solution does not depend on having defined security domains, and can defend against an attacker running on the same or another core. 

We evaluate RollingCache on ChampSim using the SPEC-2017 benchmark suite. 
Our security evaluation shows that our dynamic mapping removes the deterministic ability to identify the source of contention. The performance evaluation shows an impact of 1.67\% over a mix of workloads, with a corresponding 
area overhead of $\sim$ 5\%. 

\end{abstract}

\section{Introduction}


Computing systems are typically utilized by multiple applications at the same time. Hardware components shared across different executing entities can be used to leak information from one entity to another. Focusing specifically on the memory hierarchy, 
modern processors exploit the spatial and temporal locality properties of applications in order to improve performance. The architectural optimizations, specifically, a memory hierarchy with caches, result in non-uniformity of memory access latency. When these caches are shared across different executing entities, they potentially expose protected application information via a timing side channel, because of the difference in access times of cached and uncached data. These timing side channels have been shown capable of leaking information across process boundaries, enclave boundaries, across privilege separations, and from speculative domains~\cite{PaaS,templateattack,spectre,meltdown}.

There are two broad categories of cache side channel attacks, namely, \emph{reuse} and \emph{contention} attacks. 
\emph{Reuse} attacks exploit  shared memory (data or code) between the attacker and the victim. 
\emph{Contention} attacks depend on having a shared cache between the attacker and the victim, 
and do not require the presence of shared memory. 
In this work, we design a cache to defend against \emph{contention-based} side channel attacks, where the attacker tries to determine the victim's access pattern by identifying the cache set accessed by the victim. Since accesses to a set beyond the number of \emph{ways} in a set result in a \emph{conflict miss}, an attacker can learn if a set was accessed by incurring a conflict miss due to the victim's accesses. The many-to-one mapping between the addresses and the set they map to in the cache results in a deterministic set of addresses that may cause conflict misses in any given cache set. 

Existing defenses resort to partitioning or hide the mapping between the addresses and their location in the cache by using encryption. 
Randomizing caches either use a lookup table~\cite{newcache} to cache data at different locations for different security domains or use domain-specific encryption of address lines~\cite{mirage,ceaser,scattercache}.
Encryption and decryption of the address line~\cite{mirage, ceaser, scattercache} can take multiple extra cycles per access. Further, static encryption is susceptible to rapidly advancing attacks~\cite{probpnp}, as a set of encrypted addresses can still be constructed to launch the same attack~\cite{evictionset}. While encryption keys can be changed to prevent the discovery of eviction sets, doing so at a rate sufficient to deter the attack can result in significant area and performance impact~\cite{ceaser,ceasers}, as it is accompanied by relocating the existing data in cache. 

Another common shortfall with existing defense techniques is that they require identified security domains for the defense to be effective~\cite{scattercache,mirage,newcache, dawg}.
Unfortunately, security domains separating an attacker from the victim may not be well defined in some computing domains (for example, the attacker and victim could reside in the same process space). 



Our goal in this work is to break the many-to-one deterministic mapping of memory addresses to cache sets without the need for encryption, decryption, or relocation of the existing data in the cache, and with low area overhead. In addition, as security domains separating an attacker from the victim may not be well defined in some computing domains (for example, the attacker and victim could reside in the same process space), the design does not rely on the ability to have uniquely identified security domains.

We present RollingCache~\cite{dissertation_divya}, a cache design that utilizes 
the dynamic nature of the runtime state of the system to create the necessary nondeterminism in the addresses that contend in cache. RollingCache uses indirections for mapping a set of addresses to cache sets. 
These indirections are not static for a predetermined period, and get updated  to a new location after a certain number of cache line fills. Rather than a replacement in the same cache set when the fill limit is reached, a conflict miss updates the indirection, without requiring data relocation.
The rate at which the mapping changes is a nondeterministic function of the number of replacements in the cache. 


We avoid relocating the existing cache content by allowing lookup of additional cache sets, providing continued visibility of the data that was filled in the previous set. Addresses are thereby allowed to map to multiple cache sets, eliminating the traditional one-to-one mapping of the memory address to the cache set. 
Conversely, multiple sets of addresses may map to the same cache set. 
Combined with the continuous update of mappings at a rate that is dependent on the cache miss rate, RollingCache ensures different sets of addresses contend with and evict one another. 

The security of RollingCache relies on continuously changing mappings, making discovery of conflicting addresses difficult. 
We analyze the properties of RollingCache that prevent an attacker from deciphering access patterns by incurring conflict misses, and determine the likelihood of non-self associative evictions.

We evaluate RollingCache on ChampSim, a trace-based simulator, using SPEC2017 workloads. The traces are collected for out-of-order execution, and we use a combination of different workloads to study the impact on performance.
We use Cacti~\cite{cacti} to estimate the area overhead to be 5\%.

The key contributions made by the paper include: 
\begin{itemize} 

    \item recognizing that cache access behavior and miss rate can be used to introduce nondeterminism in mapping memory addresses to cache sets; 
    
    \item using indirection to map a set of addresses to cache sets to defend against contention-based cache side channel attacks without the need for encryption, relocation, or known security domains;

    \item updating the indirection to dynamically change the sets of addresses being contended with;
    
    \item reducing conflict misses in a given cache set and thereby enhancing performance, as a potential side effect; 


    \item providing a theoretical analysis of the security and using ChampSim to evaluate the cache performance.

 \end{itemize} 

\section{Background}
\label{sec:bg}

\subsection{Cache Access}
Caches bridge the gap between the speed of operation of CPU and memory access latency, by storing a copy of data that is likely to be accessed by the CPU. The location of data in the cache is conventionally determined by its memory address (whether physical or virtual), with the simplest approach  using index bits created by splitting the address into tag, index, and offset bits. 
More sophisticated approaches, for example, modern Intel processors, 
may use a hash function to map addresses to different slices of the last-level cache (LLC)~\cite{llcpractical}. 
The combination of slice ID and index bits from the address are then used to determine the set in which the data can be cached, resulting in a many-to-one mapping of addresses to a cache set. 

\subsection{Cache Side Channels}

There are two broad categories of cache side channels, depending on the presence of shared memory (shared code or data) between the attacker and the victim. The two types of attacks are called \emph{reuse-based} and \emph{contention-based} attacks. In the presence of shared memory, an attacker may be able to determine the address of access by a victim at a cache line size precision using a \emph{reuse-based} attack. In the absence of shared memory, the attacker can only determine the cache set accessed by the victim using 
\emph{contention-based} attacks~\cite{primeprobe}. These attacks find applicability in leaking data across virtual machines, and have been used to steal cryptographic keys from different protocols like AES and RSA~\cite{llcpractical,primeprobe}. 

Several defense techniques have been studied for preventing cache side channels. The \emph{reuse} attack is mitigated by preventing sharing~\cite{secdcp,dawg}, disabling \emph{flush} instructions, constant-time implementation~\cite{rane15raccoon}, and by TimeCache~\cite{timecache}. 
We focus in this paper on contention-based attacks.

Contention attacks exploit the static many-to-one mapping between a memory address and a cache set. This known fixed mapping allows quickly locating data in the cache. An attacker has the ability to learn the mapping over time by creating a set of addresses that evict each other from the cache~\cite{evictionset}.
 
The attacker tries to determine the cache set accessed by a victim. \emph{Prime+probe}~\cite{primeprobe} is a common contention-based attack, and it finds application in leaking cryptographic keys or learning about key strokes. To carry out the attack, the attacker needs the ability to time their own accesses. 

\subsection{Eviction Sets}
An eviction set is defined as the group of addresses that when accessed, can replace the contents of a cache set~\cite{evictionset}. In a conventional cache, addresses with the same index bits and slice ID are said to be \emph{self-associative} or \emph{congruent}. This set of self-associative addresses can be used to form eviction sets that can replace the contents in the cache set.
The notion of an eviction set is important for contention-based attacks because a subsequent slow access on one of the addresses in the eviction set implies that something else from the set of self-associative addresses was accessed.

\subsection{Hiding Eviction Sets}

Some existing solutions for contention-based attacks use encryption to hide the actual address or the set of contending addresses in the cache~\cite{mirage, ceaser, ceasers, scattercache}. This requires multiple cycles for encryption and decryption on every access. Since a static encrypted mapping can be reverse engineered given enough time, a new mapping is used every so often and the existing data in the cache is moved around~\cite{mirage, ceaser, ceasers}. Relocating existing data in cache is extra work and results in additional misses.  
Other approaches resort to domain specific encryption or indexing~\cite{scattercache, newcache}, which might not support every use case in different computing domains. 
Designs that use encryption to hide eviction sets are still susceptible to eviction set discovery over time~\cite{evictionset}. 

\section{RollingCache: Dynamic\\ Contention}
\label{sec:design}

The idea of an eviction set is important for contention-based attacks, because it leaves the attacker with a set of possible addresses to guess from, for the victim's access.
We propose a defense mechanism that renders eviction sets less informative by changing them constantly. Not only is the set of conflicting addresses unknown, it is also updated dynamically, making the discovery of the eviction set difficult.

The rate at which the set of conflicting addresses is updated is directly related to the miss rate in the conflicting address sets. This prevents the attacker from discovering the set of conflicting addresses, as the accesses for discovering the eviction set update the set of addresses that contend in cache. The addresses that can contend with each other are a function of the runtime state of the cache. 

We achieve dynamic contention by gradually moving cache occupancy of a set of addresses from one cache set to  another. 
We avoid the work involved in cryptographically segregating or moving around the content in the cache. Thus, the design of the cache is not hidden, yet the attacker cannot decipher the relation between the addresses and the cache locations.

\subsection{Contention Across Address Sets}
We identify the terms \emph{AddrSet} as the set of addresses that get indexed in to the same cache set in a conventional cache, and \emph{CacheSet} as the physical cache set location in the cache. There are as many \emph{AddrSets} as there are \emph{CacheSets}. A conventional cache has an \emph{AddrSet} mapped to a single \emph{CacheSet}.
RollingCache separates the \emph{AddrSet} from the \emph{CacheSet} and uses a level of indirection to map an \emph{AddrSet} to two \emph{CacheSets} at a time.

Repeated cache fill requests from an \emph{AddrSet} beyond a certain count, instead of causing eviction in the given \emph{CacheSet}, rolls over to fill a different \emph{CacheSet}. The data in the current filled  \emph{CacheSet} continues to be accessible while new cache fills occur in the newly assigned \emph{CacheSet}. This newly assigned \emph{CacheSet} is potentially already mapped to and filled or fillable by another \emph{AddrSet}.
This ability to contend with addresses that do not belong to the same \emph{AddrSet} is useful for preventing contention-based attacks. 

RollingCache finds a replacement victim \emph{CacheSet} over all available \emph{CacheSets}, rather than finding a replacement victim cache line within a cache set.
The above scheme allows different \emph{AddrSets} to have overlapping occupancy in a \emph{CacheSet} resulting in contention with a dynamically changing set of \emph{AddrSets}.

\subsection{Threat Model}

Cache side channel vulnerabilities exist in a range of architectures, 
and are a threat to a wide variety of computing domains, from cloud computing to end user devices, with or without the use of secure enclaves. The attacker and victim could run on different cores or the same core, or even the same process/thread of execution. Thus, the threat model for an attack can be a combination of any of the possible attack surfaces. 

RollingCache defends against contention-based cache side channel attacks in shared caches. 
We do not restrict our discussions in this work to a specific computing domain.  
We do not require any prior knowledge of specified security domains. The attacker and the victim could run on the same or different cores, and share a cache in the system. The attacker could be a single or multiple processes attacker, and could even be within the same process space. The attacker has no control over the victim's accesses, but can observe the latency of its own accesses.

\subsection{Goals and Insights}
The goal of this work is to provide an efficient defense against contention-based attacks. The defense has the following characteristics:
\begin{enumerate}
    \item utilizes run-time behavior to dynamically change the set of contending addresses in the cache, instead of spending multiple cycles for address encryption or performing periodic data relocation. 
    \item does not require predefined or identified security domains; is useful in providing defense against any attacker entity.
    \item low overhead on performance and area.
\end{enumerate}

\section{Design and Implementation}
\label{sec:imp}

\begin{figure}[tb]
\caption{RollingCache has pointers from \emph{AddrSets} to \emph{CacheSets} and a \emph{freelist} to identify \emph{CacheSets} with free cache lines.}
\label{rolldesign}
\centering
\includegraphics[width=0.35\textwidth]{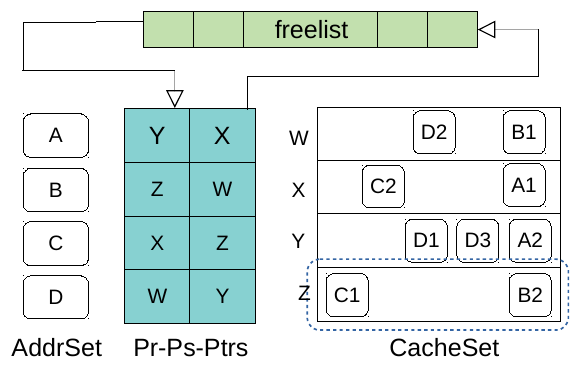}
\end{figure}

RollingCache provides dynamic mapping between \emph{AddrSets} and \emph{CacheSets}.
Figure~\ref{rolldesign} outlines the system design to show the relationship between the indirection pointers and \emph{CacheSets}, and how \emph{CacheSets} may be occupied by multiple \emph{AddrSets}. 

\subsection{Using Indirection}

The mapping of an \emph{AddrSet} to \emph{CacheSets} is available in pointers referred to as the \emph{present-ptr}(\emph{PrPtr}) and the \emph{past-ptr}(\emph{PsPtr}), and an access proceeds by identifying the mapping for the \emph{AddrSet}. The pointers are initialized randomly at the time of system initialization. The index bits and the slice ID form the \emph{AddrSet} that is used to look up the mapping for an access. Since data belonging to different \emph{AddrSets} can reside in the same \emph{CacheSet}, the \emph{AddrSet} (the index bits in a conventional cache) is stored along with the tag bits to identify the cache lines, and a tag hit further needs \emph{AddrSet} comaprison to result in a cache hit.
Both the \emph{CacheSets} pointed at by the indirection pointers of an \emph{AddrSet} may hold data for the \emph{AddrSet}. A hit can be the result of data being found in either of the two \emph{CacheSets}.   

The difference between the \emph{PrPtr} and the \emph{PsPtr} is that cache fills occur only in the \emph{CacheSet} pointed to by the \emph{PrPtr}, i.e., \emph{PrPtr[AddrSet]}, while data may be located in both \emph{PrPtr[AddrSet]} and \emph{PsPtr[AddrSet]}. In other words, the \emph{PrPtr[AddrSet]} works as the active \emph{CacheSet} for the \emph{AddrSet} and the \emph{PsPtr[AddrSet]} continues to be looked up for data that was placed in the past. 

\begin{figure}[tb]
\caption{\emph{AddrSet} A rolls from \emph{CacheSet} K to L to I}
\label{rollptr}
\centering
\includegraphics[width=0.42\textwidth]{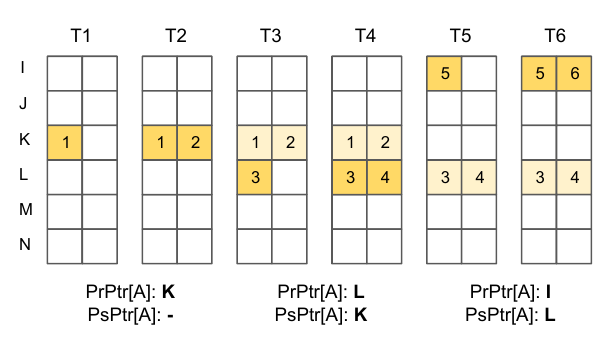}
\end{figure}

We illustrate the cache access using an example 2-way cache. Figure~\ref{rollptr} shows repeated accesses 1 through 6 to some \emph{AddrSet} A. Accesses 1, 2, are filled in \emph{CacheSet} K, because the \emph{PrPtr[A]} has the value K. Subsequent accesses 3 and 4 instead of evicting 1 or 2 fill \emph{CacheSet} L, since the \emph{PsPtr[A]} is updated to L. At this instant, the \emph{PsPtr} holds K, an access to 1 and 2 continues to hit in the cache, as the \emph{PsPtr[A]} supports lookup. Further accesses fill \emph{CacheSet} I and so on.

The dynamic nature of this mapping comes from the fact that after a certain number of replacements, the \emph{PrPtr} is updated to fill a different \emph{CacheSet} and the \emph{PsPtr} holds the previous \emph{PrPtr} to allow visibility of the recently placed data. 

\subsection{Pointer Update}
\label{ptrupdate}

 A cache lookup results in a hit if the stored \emph{AddrSet} bits in any of the cache lines in the cache sets \emph{PrPtr[AddrSet]} and \emph{PsPtr[AddrSet]} match the corresponding address bits of the cache access. Otherwise, a cache miss is incurred. A cache miss brings data into the active \emph{CacheSet}, i.e., \emph{PrPtr[AddrSet]}. After a  specific number of cache fills, a subsequent access resulting in a cache hit continues to be serviced, but a miss results in a pointer update, i.e., a miss finds a new \emph{CacheSet} in which to replace a victim cache line. We allow \emph{W} cache fills from an \emph{AddrSet} into its active \emph{CacheSet} before the pointer updates to point to a new cache set. The count is associated with every \emph{AddrSet's} \emph{PrPtr}, and is reset at a pointer update. A cache fill in a given \emph{CacheSet} takes place in one of the invalid cache lines. If there are no invalid cache lines available, a valid cache line in the \emph{CacheSet} is picked as the replacement victim at random. 

%
The current \emph{PsPtr} is saved (in a \emph{PsPtr} handling register) and replaced with the content of the \emph{PrPtr}. A replacement victim \emph{CacheSet} is picked to update the \emph{PrPtr}. The replacement victim \emph{CacheSet} is now the new active set, which gets filled with further accesses belonging to the \emph{AddrSet} under consideration. 

 Cache lines corresponding to the \emph{AddrSet} in the cache set pointed to by the saved \emph{PsPtr} must be invalidated since the \emph{CacheSet} is no longer accessible by the \emph{AddrSet} after the pointer update. The cache lines to be invalidated are identified by comparing the \emph{AddrSet} with the stored \emph{AddrSet} bits in the tag for each cache line. The \emph{PsPtr} handling register performs these invalidations and also handles any necessary writebacks. Since the time for these operations is not fixed and these operations are not atomic, misses in the cache due to accesses to a \emph{AddrSet} must check the handling register. The cache can thereby continue to operate in a non-blocking fashion as long as the writeback buffers and handling registers do not overflow. Once the invalidations and writebacks are complete, the partially freed \emph{CacheSet} is then put at the end of the \emph{freelist}. 

The sequence of operations for updating the indirection pointers are summarized as follows:
\begin{itemize}
    \item \emph{temp} = \emph{PsPtr[AddrSet]}
    \item \emph{PsPtr[AddrSet]} = \emph{PrPtr[AddrSet]}
    \item\emph{ PrPtr[AddrSet]} = replacement victim \emph{CacheSet}
    \item invalidate \emph{AddrSet} cache lines in \emph{temp} 
    \item \emph{freelist[tail]} = temp
\end{itemize}

\begin{figure}[htb]
\caption{Pointer updates result in cross \emph{AddrSet} contention}
\label{ptr2}
\centering
\includegraphics[width=0.475\textwidth]{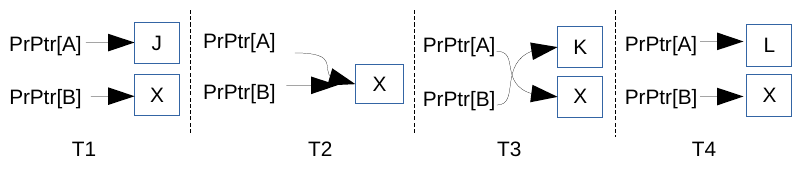}
\end{figure}

Pointer updates make it possible to have different \emph{AddrSets} conflict in a \emph{CacheSet} for the duration between pointer updates for the residing \emph{AddrSets}. This implies both \emph{self} and \emph{non-self associative} addresses can conflict in a \emph{CacheSet}. For instance, if \emph{AddrSet} A picks \emph{CacheSet} X as the replacement victim, as shown in Figure~\ref{ptr2}, which already has content from \emph{AddrSet} B, \emph{AddrSet} A and \emph{AddrSet} B can evict each other until either \emph{AddrSet} A or B undergo a pointer update and loses the ability to fill \emph{CacheSet X}. This creates cross contention between A and B, from time T2 to T3 and T2 to T4.

\subsection{Replacement CacheSet Victim}
\label{replacement}
The replacement \emph{CacheSet} victim for the pointer update is selected randomly from the \emph{freelist}, a list of \emph{CacheSets} that are highly likely to have invalid cache lines. The \emph{freelist} updates on a pointer update, i.e., an entry is removed from the list as the replacement victim, and the \emph{CacheSet} being freed is added to the end of the list. The size of the \emph{freelist} is a configurable design choice. 
The \emph{freelist} is initialized randomly with different \emph{CacheSets} at system startup.

\begin{figure}[tb]
\caption{Flowchart demonstrating pointer update. }
\label{flowchart}
\centering
\includegraphics[width=0.28\textwidth]{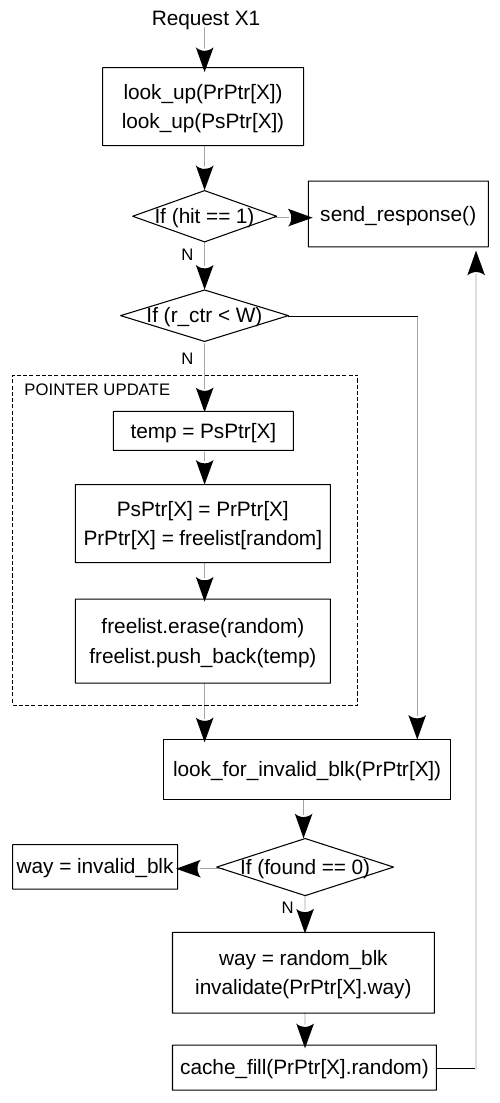}
\end{figure}

Once a new \emph{CacheSet} has been found, any invalid lines are filled before resorting to replacements, as described earlier. The entire mechanism of cache access, cache fill, and pointer update is summarized in the flowchart shown in Figure~\ref{flowchart}. The \emph{AddrSet} under consideration here for cache accesses is `X'. An access is searched in both the present and the past \emph{CacheSets}, and up to \emph{W} cache fills are allowed before the next pointer update. The pointer update identifies a replacement victim \emph{CacheSet}. Within the identified \emph{CacheSet}, a random cache line is picked for replacement in the absence of invalid lines. The pointer update latency is not on the critical path and can be hidden behind the memory access latency, since it is performed only on a miss.

The random number used to index into the \emph{freelist} can be generated from an on-chip hardware random number generator (HRNG) similar to PhantomCache~\cite{phantom}. Intel CPUs can generate random numbers at 3 Gbps~\cite{intelcpu}, which can support 500 million random indices for a \emph{freelist} with 64 entries, for instance. This is expected to be sufficient for a remap occurring after \emph{W}=16 replacements in an \emph{AddrSet}, as the average rate of replacement is lower.

\subsection{Cache Initialization}
The initial state of the cache is a randomized mapping, with a random number of writes remaining before triggering pointer updates in different \emph{AddrSets}. If the system is being attacked from the very inception, regardless of whether in kernel or user space, the \emph{freelist} is randomly initialized and the indirection pointers would start with a random mapping.

\section{Security Evaluation}
Rolling cache deters contention attacks both at the prime and the probe stage. 
Three key design features contribute to the overall effectiveness of the defense:
\begin{itemize}
    \item Unknown number of cache line fills available prior to a rollover
    \item Randomized replacement of a cache line
    \item Dynamic many-to-multiple nondeterministic mapping of address sets to cache sets
\end{itemize}

We describe how these features render an attack inconclusive in the following sections.

\subsection{Priming Sequence}
\begin{figure}[tb]
\caption{Example attack sequence for a 2-way RollingCache}
\label{toyexample}
\centering
\includegraphics[width=.47\textwidth]{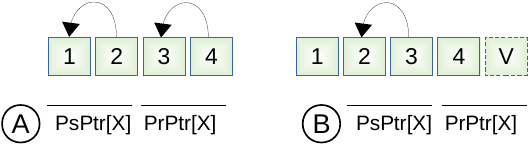}
\end{figure}

We consider an example attack sequence played under RollingCache design, with the cache being 2-way set associative. The attacker may want to make 2W accesses intending to prime the 2 CacheSets PrPtr[X] and the PsPtr[X]. Let the attacker's accesses to prime the CacheSets be to addresses 1, 2, 3, 4. The possible outcomes of priming left in the cache depends on the state of the cache, and is shown in Figure\ref{toyexample}. `A' shows the scenario where the 4 accesses start after a pointer update, B shows that priming does not start right after a pointer update. 2W accesses when not starting right after a pointer update can overflow onto a third CacheSet, losing access to the very first few accesses; the victim may access the same PrPtr[X] without evicting the primed accesses. The arrows represent that 1, 2 or 3 could be replaced by one of the later accesses during priming, as we follow random replacement within the CacheSet.

The probability that an individual access remains in the allocated CacheSet after priming can be expressed using the following equations, where the probability of scenario `A', i.e., prime begins at a pointer rollover is expressed as P(A), P(B) is the probability of scenario `B' (rollover during priming), and P($C_n$) is the probability of no random eviction of the $n$th access to a set, by any of the following accesses in the sequence, and `W' is the number of ways in a CacheSet.
\begin{align}
P(A) = & (1/W) \\
P(B) = & (W-1)/W \\
P(C_n) = & (\frac{W-1}{W})^{W-n} 
\end{align}
for W = 2,
\begin{multline}
\label{prime1}
P(1\ in\ cache\ after\ Prime) \\
= P(prime\ starts\ at\ a\ ptr\_rollover) and \\
P(no\ random\_eviction) \\
= P(A)*P(C_1) \\
= (1/W)(\frac{W-1}{W})^{W-1}            
\end{multline}
Similarly,
\begin{multline}
\label{prime2}
P(2\ in\ cache\ after\ Prime) \\
= P(prime\ starts\ at\ ptr\_rollover) or\\
P(prime doesn't\ start\ at\ ptr\_rollover)*P(no\ random\_eviction) \\
= P(A)+ P(B)*P(C_1)\\
= (1/W) + ((W-1)/W)(\frac{W-1}{W})^{W-1}\\
= (1/W) + (\frac{W-1}{W})^{W}
\end{multline}
\begin{multline}
\label{prime3}
P(3\ in\ cache\ after\ Prime) \\
= P(prime\ starts\ at\ ptr\_rollover)*P(no\ random\_ eviction)\\
or\ P(prime\ doesn't\ start\ at\ ptr\_rollover)(1)\\
= P(A)P(C_1)+ P(B)\\
= (1/W)(\frac{W-1}{W})^{W-1} + ((W-1)/W)
\end{multline}
\begin{equation}
\label{prime4}
P(4\ in\ cache\ after\ Prime) = 1
\end{equation}

The data left in the CacheSets after the prime depend on the original state of the cache. The probability of prime access 1, 2, 3, and 4 remaining in the cache at the end of the prime sequence is 0.25, 0.75, 0.75, and 1 respectively. The corresponding probability of eviction during priming and hence a miss for the probe of access $x$ due to prime ($MPrime_x$) is 0.75, 0.25, 0.25, and 0.

Incurring a miss at probe itself can trigger another pointer update, and can undo the effect of priming, which implies that the attacker may want to infer a victim's access behavior based on the very first miss at probe. However, as illustrated in Figure~\ref{toyexample}, there is no good candidate for probe, as all of 1, 2, 3 can be missing even before the victim's access due to random eviction or due to pointer rollover. 2, 3, 4 are also not good candidates for probe since a subsequent hit on these does not imply that the victim did not access the same AddrSet. They leave the possibility of the victim going undetected as shown above in Figure~\ref{toyexample},`B'. 

\subsection{Probe Inference}

Let us consider a RollingCache with `S' \emph{CacheSets}. The number of distinct \emph{AddrSets} is also `S'. Let us say that the attacker tries to determine if the victim accesses a specific \emph{AddrSet} $AddrS_i$. The first phase of the attack constitutes making 2W accesses to addresses from \emph{AddrSet} $AddrS_i$.
The attacker then allows the victim to make its access, and probes one of the addresses from the prime sequence. 

We evaluate the individual probabilities of probe accesses to each possible primed address resulting in a miss if the victim did indeed access the AddrSet of interest $AddrS_i$. 
\begin{multline}
\label{probe1}
P(Miss1\ due\ to\ AddrS_i) = P(A)P(C_1)(1) \\
= (1/W)((W-1)/W)^{W-1}
\end{multline}
Since 2 can be evicted by the victim accessing $AddrS_i$ only if the prime had started at pointer rollover
\begin{equation}
\label{probe2}
 P(Miss2\ due\ to\ AddrS_i) = P(A) = (1/W)           
\end{equation}
Since access 3 cannot be evicted due to an access from the same AddrSet, 
\begin{equation}
\label{probe3}
P(Miss3\ due\ to\ AddrS_i) = 0
\end{equation}
\begin{multline}
\label{probe4}
P(Miss4\ due\ to\ AddrS_i) \\
= P(B)*P(random\_eviction\ due\ to\ victim)\\
= ((W-1)/W)(1/W)
\end{multline}
For W=2, the above equation \ref{probe1}, \ref{probe2}, \ref{probe3} and \ref{probe4} i.e the probability of incurring a miss on a probe of prime access 1, 2, 3, and 4, given a victim access to $AddrS_i$, is 0.25, 0.5, 0, and 0.25 respectively. 

The probability of incurring a miss on a probe of access $x$ due to a victim's access of  some other \emph{AddrSet}, $AddrS_j$, depends on the probability of $x$ remaining in the cache after prime, and the probability of the \emph{CacheSet} 
being written by $AddrS_j$. Since there are $S$ \emph{CacheSets}, the probability is $1/S$ assuming a uniform and random mapping distribution.  
The attacker’s probe access resulting in a miss is likely to coincide with the location of a victim’s access with a probability 1/W.

\begin{multline}
\label{miss1sj}
P(Miss1\ due\ to\ AddrS_j)=(P(1 in cache after Prime)\frac{1}{SW}\\
=[\frac{1}{W}*(\frac{W-1}{W})^{W-1}]\frac{1}{SW}
\end{multline}

\begin{multline}
\label{miss2sj}
P(Miss2\ due\ to\ AddrS_j)=(P(2 in cache after Prime))\frac{1}{SW}\\
=[\frac{1}{W}+\frac{W-1}{W}*(\frac{W-1}{W})^{W-1}]\frac{1}{SW}
\end{multline}

\begin{multline}
\label{miss3sj}
P(Miss3\ due\ to\ AddrS_j)=(P(3 in cache after Prime))\frac{1}{SW}\\
=[\frac{1}{W}*(\frac{W-1}{W})^{W-1}+\frac{W-1}{W}]\frac{1}{SW}
\end{multline}

\begin{multline}
\label{miss4sj}
P(Miss4\ due\ to\ AddrS_j)=(P(4 in cache after Prime))\frac{1}{SW}\\
=\frac{1}{SW}
\end{multline}

We can calculate the probability of a probe incurring a miss on the $x$th access in the prime sequence due to the victim's access to any AddrSet in the cache as follows: 
\begin{multline}
\label{probemiss}
P(M_x\ due\ to\ victim) \\
= P(accessing\ AddrS_i)*P(M_x\ due\ to\ AddrS_i) +\\ 
 P(accessing\ any\ other\ AddrS_j)*P(M_x\ due\ to\ AddrS_j)
\end{multline}
Since all the S sets are equally likely, the probability of the access being from $AddrS_i$ or $AddrS_j$ is 1/S and there are (S-1) other AddrSets besides $AddrS_i$.


Using equations \ref{miss1sj}, \ref{miss2sj}, \ref{miss3sj}, \ref{miss4sj} and for S=64 (and W=2), the probability of getting a miss due to the victim's access to any AddrSet in equation\ref{probemiss} is 0.0058, 0.0136, 0.0058, and 0.0116 respectively on a probe access to the first, second, third, and fourth prime accesses respectively.

\emph{\textbf{To find:~}} The probability of the victim's access being from \emph{AddrSet} $AddrS_i$, given a miss is incurred by the attacker during probe, i.e.,  P($AddrS_i|M_x$).\\
P($AddrS_i$) is the probability of the victim's access being from \emph{AddrSet} $AddrS_i$. 

The required conditional probability can be determined using Bayes' theorem~\cite{bayes}, which describes the probability of an event in terms of prior knowledge of conditions related to the event. The probability of event $A_i$ occurring given B is true is given as below.
\begin{align}
\label{bayes}
  P(A_i|B)= \frac{P(B|A_i)P(A_i)}{P(B)} 
\end{align}
Assuming a uniform distribution, the probability of an access being from \emph{AddrSet} $AddrS_i$ is 1/S.  
\begin{align}
\label{eq3}
   P(AddrS_i) = 1/S. 
\end{align}
Similarly,
\begin{align}
\label{eq4}
   P(AddrS_j) = 1/S. 
\end{align}
Using~\ref{bayes}, we can express the P($AddrS_i|M_x$) in terms of the probability of a miss occurring due to the victim accessing different \emph{AddrSets} or due to an eviction during priming (miss due to prime ($MPrime_x$)) as follows:
\begin{align}
\label{eq2}
  P(\frac{AddrS_i}{M_x})= \frac{P(\frac{M_x}{AddrS_i})P(AddrS_i)}{P(MPrime_x) + P(M_x\ due\ to\ victim)} 
\end{align}

\begin{itemize}
\item For S=64 (and W=2), the probability of a miss due to the victim having accessed $AddrS_i$ given a miss on a probe of the first, second, third, and fourth prime accesses is 0.0052, 0.0299, 0, and 0.3368 respectively. This probability is expected to be 1 in conventional cache.
\item The corresponding probabilities of miss due to victim having accessed a different AddrSet given a miss at probe are similar, i.e 0.0025, 0.0217, 0.0227, 0.6632 respectively. This probability for a conventional cache is expected to be 0. The probability is low for the first three prime accesses because of the high probability of self-eviction at prime ($P(MPrime_x)$ is high). Hence, this makes it hard for a probe of the third access to infer that the miss was due to the victim. 
\item Additionally, based on P($Miss_x\ due\ to\ AddrS_i$), 75\%, 50\%, 100\%, and 75\% of the time, the access to $AddrS_i$ by the victim will go undetected due to a lack of a miss on the corresponding probe access, further reducing the probability of information leak for all probe accesses, including that for a probe of the fourth prime access. 
\end{itemize}

%


In summary, RollingCache's design introduces nondeterministic behavior at the prime, victim, and probe accesses, resulting in high error rates in victim access prediction and making iterations of attack to decipher the victim's access ineffective. Unlike in a conventional cache, inferring the AddrSet accessed from the misses is not deterministic due to contention spread across different AddrSets. Further, repeating the attack will not see the same behavior, as the contention for a second run of the attack might be in other CacheSets with different states from the first run.

\subsection{Freelist}

The \emph{freelist} is a dynamic structure that allows rotation of \emph{CacheSets} across different \emph{AddrSets}. Its entries are as dynamic as the \emph{AddrSets} that are accessed and undergo pointer update. 
The length of the \emph{freelist} cannot be manipulated by the attacker, as every change to the \emph{freelist} removes an entry and puts an entry on the list, resulting in no net effect on the list length. This means that the attacker cannot reduce the length of the \emph{freelist} to make its behavior deterministic.
Further, the entries on the \emph{freelist} are a microarchitectural state not visible to the attacker: they change at pointer update, i.e. at instants unknown to the attacker. The attacker is not in control of which entries can be put on the \emph{freelist} and at which time, since it depends on the prior metadata.

\begin{figure}[tb]
\caption{Accessing a specific \emph{AddrSet} allows \emph{CacheSets} to rotate through the \emph{freelist} but does not change the existing mapping of other \emph{AddrSets} to the same \emph{CacheSets}.}
\label{freelist}
\centering
\includegraphics[width=0.21\textwidth]{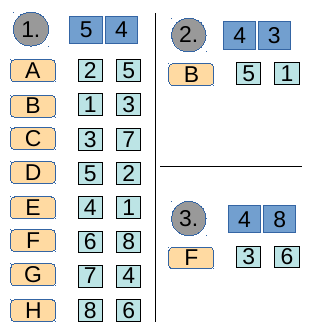}
\end{figure}

An attacker attempting to access a specific \emph{AddrSet} constantly will cause the data from that \emph{AddrSet} to roll over to \emph{CacheSets} that may be picked from the \emph{freelist} on pointer updates. A subsequent access by the victim to any of the other \emph{AddrSets} that were potentially mapped to any of these \emph{CacheSets} prior to the attacker's accesses, may result in a miss. Since these potentially contending \emph{AddrSets} are not known to the attacker, they are unable to disambiguate among the \emph{AddrSets} accessed by the victim. Figure~\ref{freelist} shows a cache with 8 \emph{CacheSets} and a \emph{freelist} of length 2. We discuss a scenario of different pointer updates to demonstrate the effect on an attack. A through H are the \emph{AddrSets}, the mapping pointers are indicated alongside the \emph{AddrSets}, with the \emph{freelist} entries listed at the top of each scenario.
\begin{enumerate}[]
    \item shows random initialization for the mappings and the \emph{freelist}. Accesses occur from different \emph{AddrSets} and their mapping undergoes an update after W cache fills.
    \item shows the instant at which \emph{AddrSet} B undergoes a pointer update, \emph{CacheSet} 5 is picked as the replacement victim, and 3 is added to the \emph{freelist}.
    \item subsequently, the \emph{AddrSet} F undergoes a pointer update, picks \emph{CacheSet} 3, and puts 8 back on the list.
    \item if the attacker has exclusive access over the cache and is interested in \emph{AddrSet} F, it makes several accesses to \emph{AddrSet} F, effectively moving data between \emph{CacheSets} 3,6,4,8, with any two of these \emph{CacheSets} being accessible at any time.
    \item a subsequent delayed access after a victim's access could be due to contention from \emph{AddrSet} C, E, H, or F (in \emph{CacheSets} 3, 4, 8, 6). We can infer this given the mappings. The attacker however, has no knowledge of the microarchitectural state of the cache, so does not know the smaller set of potentially conflicting \emph{AddrSets} at this instant.
\end{enumerate}
The above illustrates that the attacker's repetitive accesses to the same \emph{AddrSet} does not change the existing mapping of other \emph{AddrSets}. Subsequent conflict misses from the other \emph{AddrSets} mapping to the same \emph{CacheSets} interfere with the attack.

\subsection{AddrSet Cache Footprint}

\begin{figure*}[t!]
    \centering
    \begin{subfigure}[b]{0.48\textwidth}
        \centering
        \includegraphics[width=.98\linewidth]{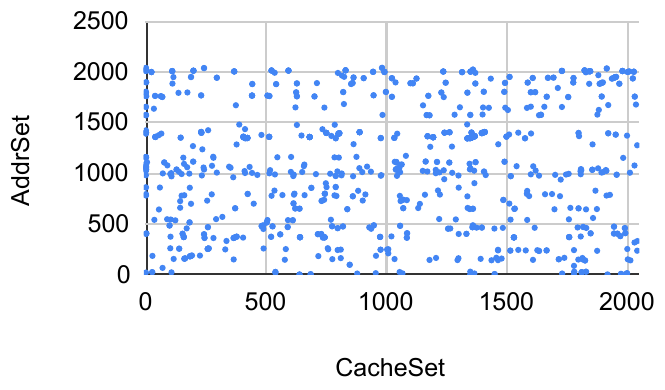}
        \caption{Smaller sample showing multiple AddrSet occupancy}
        \label{fprint_graph}
    \end{subfigure}%
    ~ 
    \begin{subfigure}[b]{0.48\textwidth}
        \centering
        \includegraphics[width=.98\linewidth]{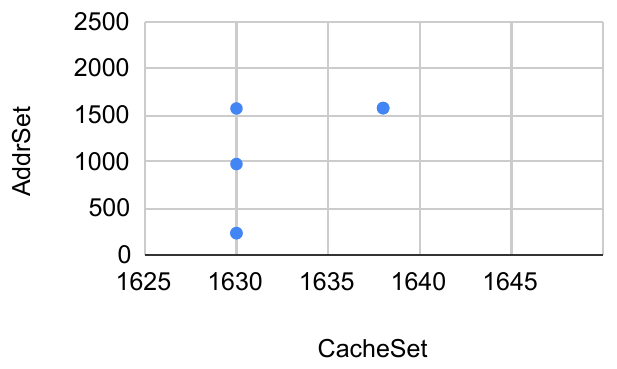}
        \caption{Closer view of the multiple AddrSets mapped to a CacheSet}
        \label{zoom}
    \end{subfigure}
    \caption{Present-Ptr footprint of different \emph{AddrSets}}
    \label{perf}
\end{figure*}

Figure~\ref{fprint_graph} shows the mapping of different \emph{AddrSets} to \emph{CacheSets} using the \emph{PrPtr}, over 1000 LLC accesses in a sample application (pop2), run with 10 million warmup instructions.  The dots represent the different \emph{AddrSets} on the y-axis that occupy the \emph{CacheSets} on the x-axis. The size of the LLC is 2MB, and it has 2048 \emph{CacheSets}. 
The scatter plot demonstrates how accesses to an \emph{AddrSet} are spread across different \emph{CacheSets}. Figure~\ref{zoom} shows a fewer \emph{AddrSets} and their active \emph{CacheSets}. A closer look into the snapshot shows that a single \emph{CacheSet} 1630 is the active \emph{CacheSet} for 3 different \emph{AddrSets} within the 1000 accesses window.

\section{Performance Evaluation}
\label{sec:eval}

\begin{figure*}[tb]
\caption{Execution time normalized to the baseline for SPEC2017 benchmarks run on 1 core; LLC size 2MB}
\label{perf}
\centering
\includegraphics[width=.9\textwidth]{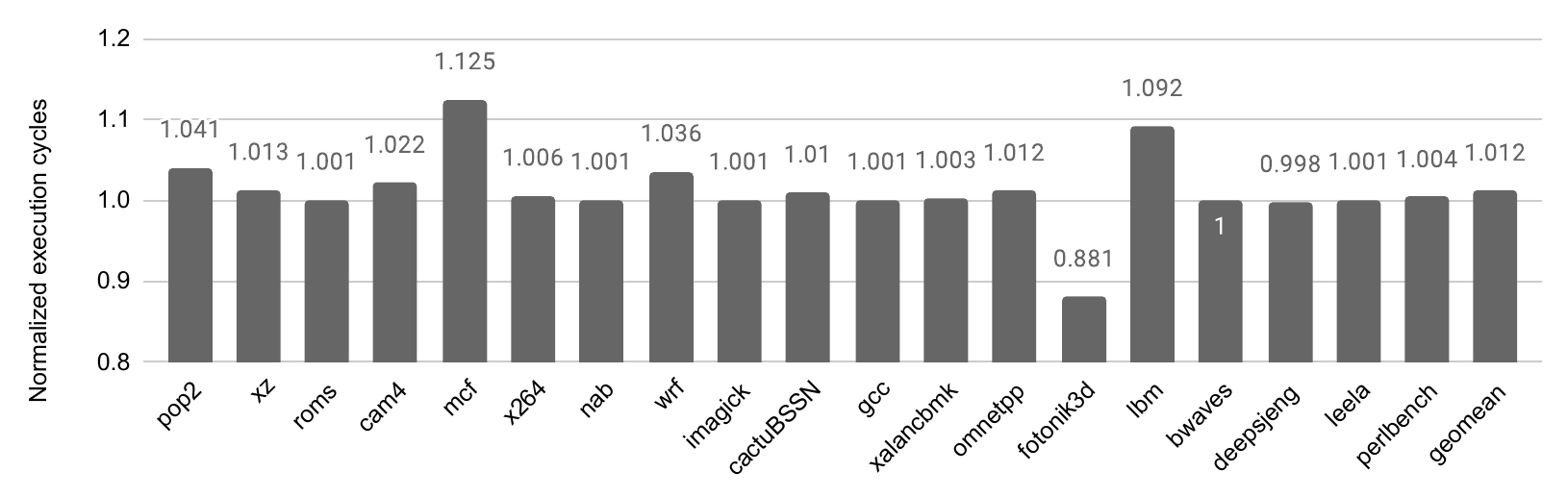}
\end{figure*}

\begin{figure}[tb]
\caption{LLC average miss latency normalized to the baseline }
\label{lat}
\centering
\includegraphics[width=0.45\textwidth]{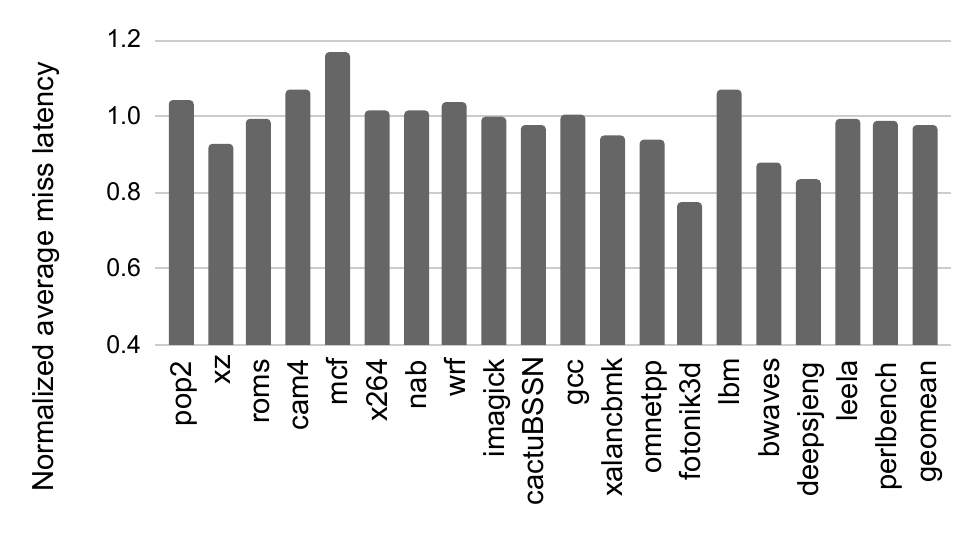}
\end{figure}

\begin{figure*}[t!]
    \centering
    \begin{subfigure}[b]{0.48\textwidth}
        \centering
        \includegraphics[width=.98\linewidth]{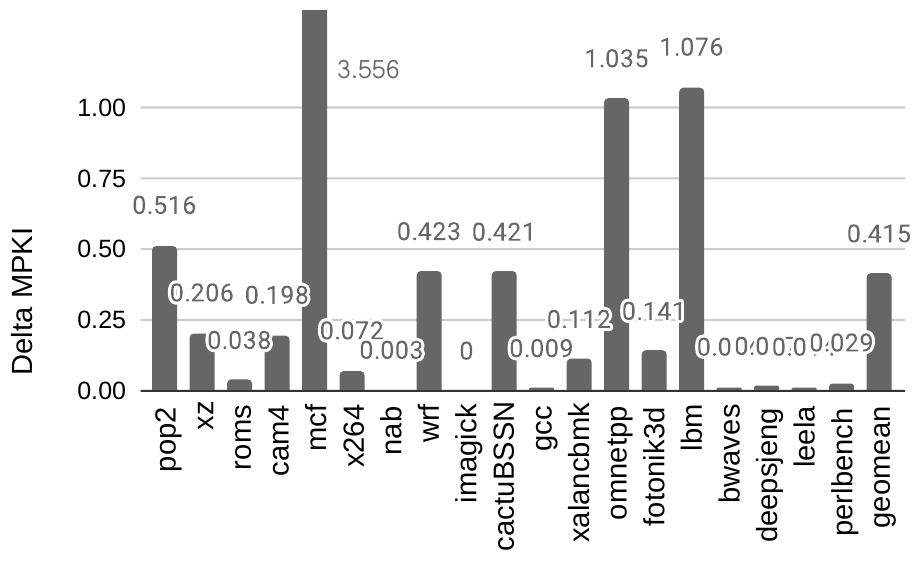}
        \caption{Change in LLC MPKI relative to the baseline}
        \label{deltampki}
    \end{subfigure}%
    ~ 
    \begin{subfigure}[b]{0.48\textwidth}
        \centering
        \includegraphics[width=.98\linewidth]{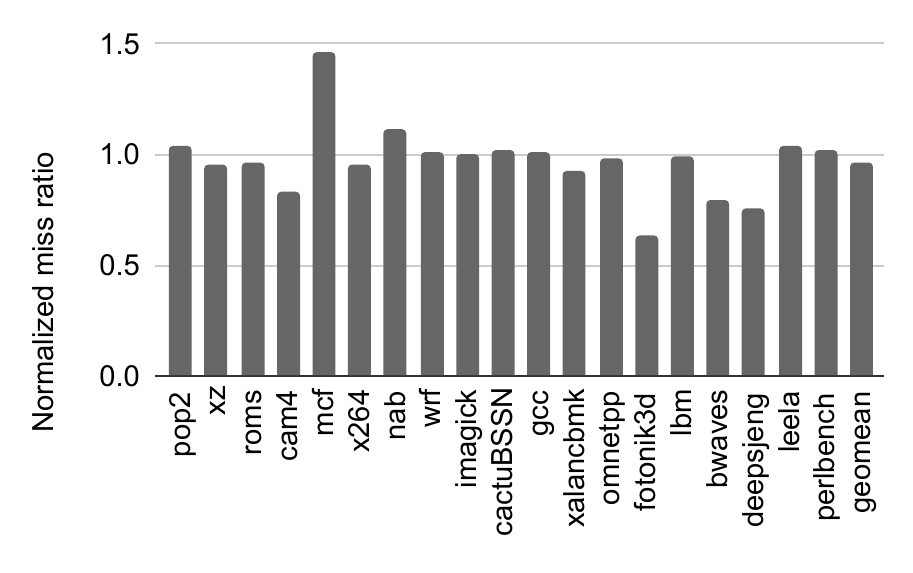}
        \caption{Rowbuffer miss ratio normalized to the baseline, for read requests to DRAM}
        \label{rq}
    \end{subfigure}
    \caption{The increase in performance in some cases, despite increased number of misses at LLC correlates with reduced miss ratio at read queue}
    \label{perf}
\end{figure*}

\begin{figure*}[tb]
\caption{IPC and misses normalized to baseline for a mix of benchmarks from SPEC2017; each mix is 4 different workloads running on 4 cores, sharing the LLC of size 8MB.}
\label{perf2}
\centering
\includegraphics[width=0.9\textwidth]{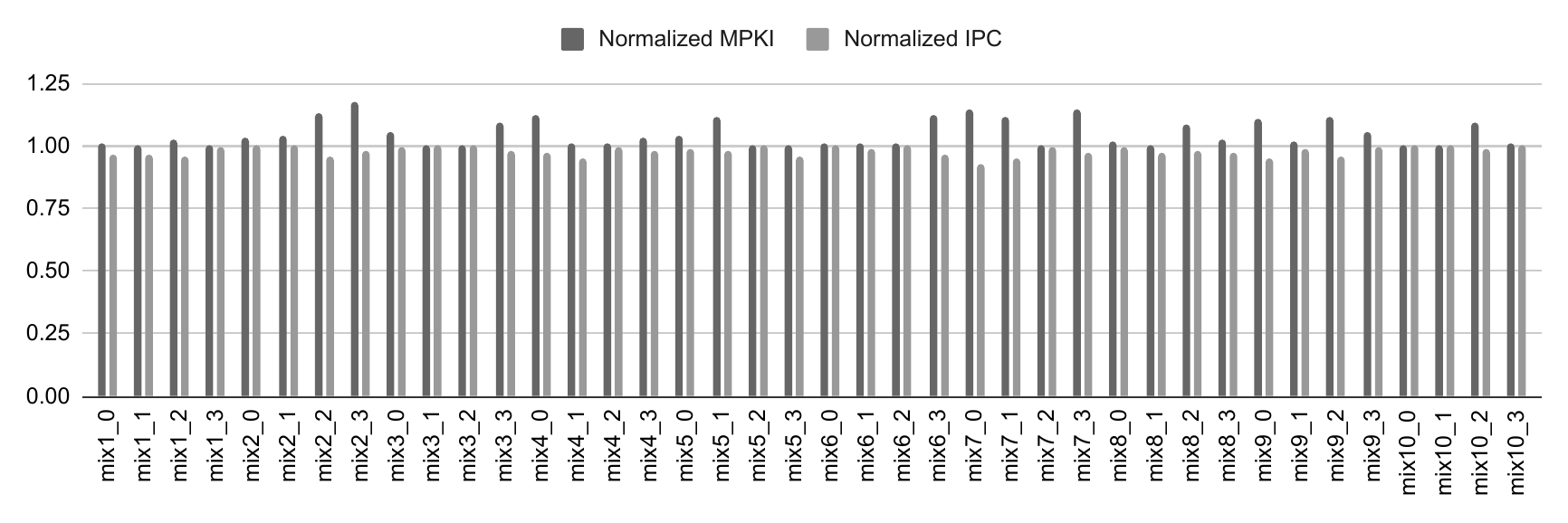}
\end{figure*}

We evaluate performance using ChampSim, which is a trace-based cache simulator~\cite{champsim}, using SPEC2017 benchmark traces for out-of-order execution. We study the impact on misses and the runtime for both single and a mix of benchmarks. The cache configuration used for the evaluation is listed in Table~\ref{cachetable}. We run a mix of 4 different workloads on 4 cores, with LLC size 8MB, as well as single benchmarks on a single core with LLC size 2MB. Our evaluation is conducted only at the LLC, but the design is not specific to a cache level and is suitable for any level of cache. The baseline used for the comparison is a conventional LRU cache of the same size. 
We evaluate the RollingCache design with a \emph{freelist} of size 64. An evaluation with 32 entries showed similar performance. A bigger \emph{freelist} provides a larger pool of cache sets within which a given \emph{AddrSet} can contend with other \emph{AddrSets}.

The single benchmark instance is run on a single core for 1 billion instructions with an initial warmup of 10 million instruction. The average increase in the misses per thousand instruction (MPKI) is 0.415, and the slowdown due to the change in the number of misses is about 1.15\% with respect to the baseline. 
RollingCache's replacement policy performs better in comparison to a random set associative replacement policy, which incurs an increase in MPKI of 0.93 and increases normalized execution time by 1.38\%.
The additional cost of looking up the indirection pointers for RollingCache with 8K sets can add up to 1 extra cycle (refer Section~\ref{cacti}), effectively making the overhead 1.34\%.

Figure~\ref{perf} shows the impact of RollingCache on the execution time of individual benchmarks. Relative to the baseline LRU, fotonik3d shows a 12\% increase in performance, which can be seen to be correlated with a reduction in misses at the DRAM row buffer for read requests, as shown in Figure~\ref{rq}.  
Conversely, mcf shows a 12\% reduction in performance, which can be correlated with the increase in MPKI, as shown in Figure~\ref{deltampki}. 
Note, however, that the increase in MPKI is not directly correlated with a decrease in performance or with the baseline MPKI (as can be seen by comparing the MPKI and normalized execution time of omnetpp, lbm, and wrf).
Rather, it is a function of the interaction of the application's access patterns with the replacement policy, as well as its ability to tolerate the access latency. 
Overall, the net effect or the average slowdown due to this replacement policy on different benchmarks is low, and the policy provides defense against side channel leaks. The increased row buffer hits for read requests in some applications are potentially from accessing recently written back data. Even with increased number of misses, the hits at the row buffer reduce average LLC miss latency, and result in increased performance. The reduction in the LLC miss latency is shown in Figure~\ref{lat}. 

Table~\ref{table:perftable} lists the exact miss ratios and MPKI for the baseline set-associative cache with LRU replacement, set-associative with random replacement, and RollingCache. RollingCache MPKI is lower than that of set-associative with random replacement policy due to RollingCache's side effect of increased associativity. 

We also evaluate a mix of benchmarks on 4 cores for 500 million instructions, with a warmup of 10 million instructions. Figure~\ref{perf2} shows the overheads for the mix of workloads in terms of instructions per cycle (IPC) and MPKI normalized to baseline. 

\begin{table}[h!]
  \centering
\caption{Baseline cache configuration}
  \label{cachetable}
\begin{tabular}{ |c|c| } 
 \hline
 Cache line & 64 Bytes\\
 \hline
 L1I & 8-way, 32 KB, latency-4 cycles\\
 L1D & 12-way, 48 KB, latency-5 cycles\\
 L2 & 8-way, 512KB, latency-10 cycles\\
 \hline
 LLC & 16-way, 2MB per core, latency-20 cycles\\
 \hline
 DRAM & 1 channel, 4GB\\
 \hline
\end{tabular}
\end{table}

\begin{scriptsize}
\begin{table}[h!]
	\fontsize{8pt}{8pt}\selectfont
  \centering
\caption{SPEC2017 execution time overhead, 2MB LLC MPKI }
  \label{table:perftable}
\begin{tabular}{ |c|c|c|c|c|c|c| } 
 \hline
 Workload & \thead{LRU\\MPKI} & \thead{LRU\\MR}& \thead{Rand.\\MPKI} & \thead{Rand.\\MR}& \thead{RC\\MPKI}& \thead{RC\\MR}\\
 \hline
 \hline
pop2&3.487&0.37&4.218&0.45&4.019&0.43\\
xz&1.103&0.38&1.382&0.47&1.314&0.45\\
roms&1.988&0.72&2.122&0.76&2.028&0.73\\
cam4&5.138&0.49&6.477&0.62&5.359&0.51\\
mcf&11.958&0.41&14.977&0.52&15.551&0.54\\
x264&1.137&0.63&1.237&0.68&1.206&0.66\\
nab&0.195&0.94&0.197&0.95&0.198&0.95\\
wrf&5.075&0.56&5.668&0.63&5.508&0.61\\
imagick&0.04&0.99&0.04&0.99&0.04&0.99\\
cactuBSSN&7.471&0.70&8.137&0.76&7.898&0.74\\
gcc&8.506&0.99&8.52&0.99&8.515&0.99\\
xalancbmk&1.28&0.31&1.462&0.36&1.398&0.34\\
omnetpp&9.146&0.50&10.799&0.59&10.207&0.56\\
fotonik3d&8.765&0.50&10.485&0.60&8.943&0.51\\
lbm&35.773&0.56&42.90&0.68&36.953&0.58\\
bwaves&0.206&0.56&0.243&0.65&0.221&0.59\\
deepsjeng&0.263&0.56&0.303&0.64&0.28&0.59\\
leela&0.12&0.54&0.116&0.52&0.135&0.60\\
perlbench&0.333&0.63&0.343&0.65&0.362&0.69\\
\hline
average&5.368&0.596&6.296&0.66&5.797&0.64\\
\hline
\end{tabular}
\end{table}
\end{scriptsize}

\subsection{Space and Indirection-Lookup Overhead}
\label{cacti}

We use Cacti6.0~\cite{cacti} to estimate the area overhead of the design assuming a 32nm technology. The area overhead is primarily from the additional bits required to store the \emph{AddrSet} along with the tag. This requires 14 additional bits per cache line for an 8MB LLC with 8192 cache sets.
Since we do not use LRU for finding the replacement victim, we can reduce the LRU bits used for each cache line. Assuming 4 bits for LRU in a 16-way cache, this reduces the effective additional bits to 10 per cache line, i.e., 1.85\%. 

Additional area overheads are from the indirection pointers per address set. An 8MB LLC with 8192 cache sets requires 14 bits for each indirection pointer. These pointers are per \emph{AddrSet} and not per cache line and the counter for the number of cache fills allowed before a pointer update requires 4 bits for a 16-way cache, totalling to 0.34\% . 
The \emph{freelist} and the logic implementation need a few thousand additional gates. 

The indirection can be implemented as a direct-mapped cache, with the index bits used to look up the indirection pointers. A 16-way, 64 byte block cache of 8MB requires 8k entries in the indirection table, with each indirection being 14 bits. The access time using Cacti for a direct-mapped cache with 8K blocks of 4 bytes (for 2 indirections) is estimated to be 0.203ns, which on a 32nm technology processor with up to 3.4 GHz processor is within 1 cycle. Once the set pointers are identified, two sets of tags must be read in order to determine if there is a hit, requiring multi-porting the tag array and adding 2.89\% to the area overhead. We make the design choice to access the tag and data array sequentially to keep the area overheads low by not multi-porting the data array. Once the tag has been found in one of the two \emph{CacheSets}, the corresponding cache line can be read from the data array. Sequentially accessing the multi-ported tag array is estimated using Cacti to incur an additional 2 cycles of delay. 

The overall increase in area for an 8MB cache is 5\%. The performance slowdown with a total of 3 cycles delay for the multi-ported sequential tag access is found to be 1.67\%. In comparison, encryption-based defense techniques require 3 to 5 additional cycles for encryption or decryption of the physical address associated with any cache access~\cite{mirage, scattercache}.


\subsection{LLC Size Sensitivity Analysis}

\begin{figure}[tb]
\caption{Performance sensitivity to larger caches sizes }
\label{sens}
\centering
\includegraphics[width=0.43\textwidth]{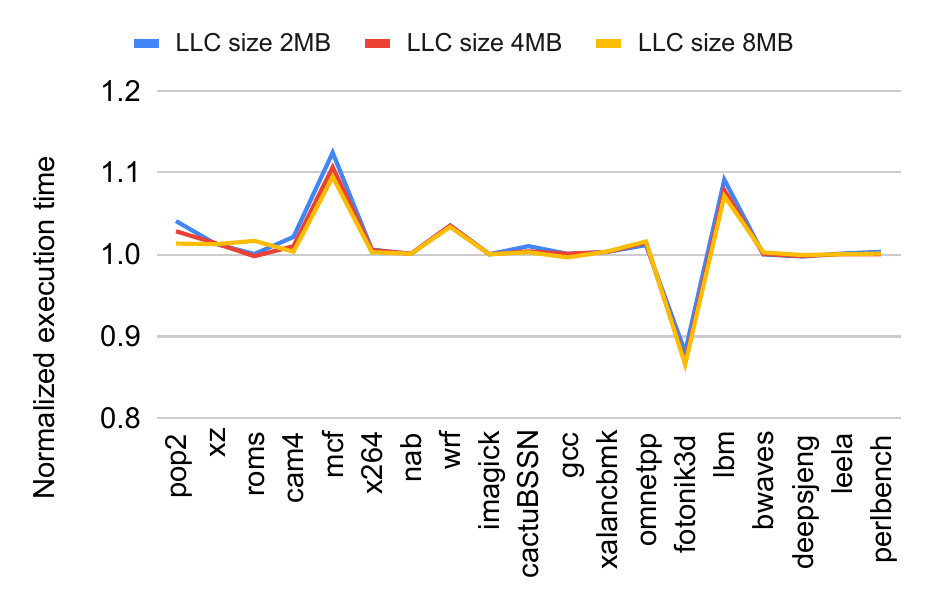}
\end{figure}

We evaluate the performance with increasing size of the LLC by changing the size of the LLC from 2MB to 4 and 8MB per core. Figure~\ref{sens} shows that the overheads for the bigger caches are similar to that of the smaller cache. Thus the design is not sensitive to the cache size and is suitable for large caches as well. 

\section{Other Contention Attacks}
\label{sec:other}

\subsection{LRU Attack}
LRU attacks also rely on having a group of addresses that replace or contend in the same cache set. The attacker accesses W-1 lines, allows the victim to execute, and then accesses another line. Depending on whether the attacker's W-1 lines got evicted, the attacker gets the information of whether the victim accessed its already existing cache line. RollingCache would prevent this form of attack both because it does not use an LRU replacement policy.

\subsection{Evict + Time}

Another attack proceeds by evicting cache sets between victim's execution, by accessing {\em W} addresses corresponding to the cache sets.
Depending on whether the second execution by the victim took longer, the attacker knows that the victim accessed a particular cache set that the attacker filled with its accesses~\cite{primeprobe}. This attack is called \emph{evict+time} and is more prone to noise than \emph{prime+probe}.

Since \emph{evict+time} also utilizes an eviction set, the attack would not work on RollingCache. Even if the attacker evicts some data corresponding to an \emph{AddrSet} by accessing more than the required number of entries from the same address set, a subsequent delay might be due to an accesses to non-self-associative addresses.

\section{Related Work}
\label{sec:related}


\subsection{Constant Time Implementation}
Some methods have been proposed to provide alternate timing primitives with reduced clock resolution~\cite{fantastic}. 
Other approaches suggest program transformation for constant time implementation~\cite{rane15raccoon,rane16,ghostrider}. These program transformations have impractical overhead due to making each critical access {\em O(n)}~\cite{rane15raccoon} and are not useful for regular applications. 

\subsection{Cache Partitioning}

Catalyst.~\cite{catalyst} and Apparition~\cite{dong2018shielding} use Intel Cache Allocation Technology (CAT) to achieve cache partitioning. Catalyst provides security for identified secure pages of memory while Apparition does so for each application. Apparition flushes the cache partition across context switches.
The overhead for Apparition varies from being negligible to about 81\% or more~\cite{dong2018shielding}. 
DAWG~\cite{dawg} partitions cache by ways similar to CAT. It restricts both hits and replacements to the ways allocated to a protection domain and allows up to 16 different domains at a time. 
The partitioned locked cache (PLCache)~\cite{newcache} modifies the cache design to allow a security domain to lock cache line. 

Partitioning reduces the effective size of the usable cache for an application, potentially reducing overall performance. 
RollingCache allows entire cache access to all applications as it does not resort to partitioning.


\subsection{Domain-Specific Lookup}
NewCache and Random Permute cache (RPcache) randomize memory mapping for different security domains using lookup tables. 
RPCache uses a per security domain \emph{permutation table} (PT)~\cite{newcache} to map set index bits to a new set index. Randomization is achieved by periodically swapping PT entries, with corresponding data movement as needed. 
The space overhead for storing permute tables for multiple security domains and large caches can be high, and hence this technique is suitable only for smaller caches. 
Instead of remapping sets, the lookup table in NewCache~\cite{newcache-micro-2016} is accessed as a direct-mapped cache and the content of the lookup table holds the cache line number. 

Both partitioning and domain-specific lookup provide side channel defense only across protection domains.  RollingCache does not require the knowledge of security domains, and is therefore able to defend against fast evolving attacks and varied threat models.

\subsection{Encryption}
Some defense techniques hide eviction sets by encrypting memory addresses. Ceaser~\cite{ceaser} encrypts addresses with a new key every epoch and relocates cache content in a time that is lower than the time required to launch an attack. With advanced forms of attack, the remap rate can get high enough to be inefficient. 
Ceaser-S~\cite{ceasers} uses encryption on a skewed cache to bring down the remap rate, while ScatterCache~\cite{scattercache} uses domain-specific encryption. 
Encryption adds significant area and power overheads and has been shown to be vulnerable to more recent forms of attacks~\cite{probpnp}.

\subsection{Increased Associativity}
Mirage~\cite{mirage} adds skewed associativity and encrypted tags to a v-way cache~\cite{vway}. A v-way cache uses indirection between tags and data lines, and provides extra tags (relative to the number of ways) in each cache set, resulting in potentially increased associativity. The total data storage capacity remains unchanged, which means that there are invalid tags at all times. Mirage uses static domain-specific encryption and supports up to 256 security domains. Static encryption has been shown to be vulnerable to eviction set discovery earlier~\cite{probpnp}. It also incurs a significant area and power overhead of $\sim$20\%, where the area overhead of RollingCache is only 3.12\%.

PhantomCache~\cite{phantom} places an incoming line in one of 8 sets and searches for data by looking up all 8 sets. This effectively increases the eviction set size. There remains a fixed set of lines that contend with each other and might be used by the attacker to leak information. PhantomCache consumes 67\% more energy for looking up 8 ways.

HybCache~\cite{hybcache} provides protection only for something already identified as security critical. It allows fully associative cache access over some ways in every cache set. The security critical accesses get cached in the fully associative subset of the cache, which means access to a much smaller cache. Workloads that are not security critical use the entire cache. The design is not suited to large caches.

RollingCache allows cache lookup in two cache sets, while cache fills are allowed in only 1 cache set. Allowing lookup in multiple cache sets reduces the need to relocate cache lines that are already present in the cache. Dynamically changing the address set to cache set mapping using runtime behavior eliminates the ability to determine the eviction set, while also eliminating the need for encryption, and consequently, the area overheads of encryption. 

\section{Conclusion}
\label{sec:conclusion}

We implement and evaluate RollingCache to defend against contention-based side channel attacks in caches. The defense provides dynamic contention over different address sets rather than hiding eviction sets through encryption. We decouple address sets from cache sets, and use indirection to point accesses from address sets to cache sets. Address sets are allowed a specified number of fills to a cache set, beyond which the mapping is updated, and the address set continues to fill a different cache set. We do not require data relocation on a pointer update as we allow cache lookup in the cache set filled previously. The dynamically updating indirection pointers allow contention between different address sets at different instances of time. 
Performance evaluation on ChampSim shows an average slowdown of 1.67\%, and the area overhead incurred is 5\%.

  \section*{Acknowledgment}
This work was supported in part by NSF grant award CNS-1900803.


\bibliographystyle{abbrv}
\bibliography{roll}

\begin{thebibliography}{10}

\bibitem{intelcpu}
\url{https://www.intel.com/content/www/us/en/developer/articles/guide/intel-digital-random-number-generator-drng-software-implementation-guide.html}.

\bibitem{champsim}
\url{https://github.com/ChampSim/ChampSim}.

\bibitem{hybcache}
G.~Dessouky, T.~Frassetto, and A.-R. Sadeghi.
\newblock Hybcache: Hybrid side-channel-resilient caches for trusted execution
  environments.
\newblock In {\em 29th $\{$USENIX$\}$ Security Symposium ($\{$USENIX$\}$
  Security 20)}, pages 451--468, 2020.

\bibitem{dong2018shielding}
X.~Dong, Z.~Shen, J.~Criswell, A.~L. Cox, and S.~Dwarkadas.
\newblock Shielding software from privileged side-channel attacks.
\newblock In {\em 27th $\{$USENIX$\}$ Security Symposium ($\{$USENIX$\}$
  Security 18)}, pages 1441--1458, 2018.

\bibitem{templateattack}
D.~Gruss, R.~Spreitzer, and S.~Mangard.
\newblock Cache template attacks: Automating attacks on inclusive last-level
  caches.
\newblock In {\em 24th $\{$USENIX$\}$ Security Symposium ($\{$USENIX$\}$
  Security 15)}, pages 897--912, 2015.

\bibitem{dawg}
V.~Kiriansky, I.~Lebedev, S.~Amarasinghe, S.~Devadas, and J.~Emer.
\newblock Dawg: A defense against cache timing attacks in speculative execution
  processors.
\newblock In {\em 2018 51st Annual IEEE/ACM International Symposium on
  Microarchitecture (MICRO)}, pages 974--987. IEEE, 2018.

\bibitem{spectre}
P.~Kocher, D.~Genkin, D.~Gruss, W.~Haas, M.~Hamburg, M.~Lipp, S.~Mangard,
  T.~Prescher, M.~Schwarz, and Y.~Yarom.
\newblock Spectre attacks: Exploiting speculative execution.
\newblock {\em arXiv preprint arXiv:1801.01203}, 2018.

\bibitem{meltdown}
M.~Lipp, M.~Schwarz, D.~Gruss, T.~Prescher, W.~Haas, S.~Mangard, P.~Kocher,
  D.~Genkin, Y.~Yarom, and M.~Hamburg.
\newblock Meltdown.
\newblock {\em arXiv preprint arXiv:1801.01207}, 2018.

\bibitem{ghostrider}
C.~Liu, A.~Harris, M.~Maas, M.~Hicks, M.~Tiwari, and E.~Shi.
\newblock Ghostrider: A hardware-software system for memory trace oblivious
  computation.
\newblock {\em ACM SIGPLAN Notices}, 50(4):87--101, 2015.

\bibitem{catalyst}
F.~Liu, Q.~Ge, Y.~Yarom, F.~Mckeen, C.~Rozas, G.~Heiser, and R.~B. Lee.
\newblock Catalyst: Defeating last-level cache side channel attacks in cloud
  computing.
\newblock In {\em 2016 IEEE international symposium on high performance
  computer architecture (HPCA)}, pages 406--418. IEEE, 2016.

\bibitem{newcache-micro-2016}
F.~{Liu}, H.~{Wu}, K.~{Mai}, and R.~B. {Lee}.
\newblock Newcache: Secure cache architecture thwarting cache side-channel
  attacks.
\newblock {\em IEEE Micro}, 36(5):8--16, Sep. 2016.

\bibitem{llcpractical}
F.~Liu, Y.~Yarom, Q.~Ge, G.~Heiser, and R.~B. Lee.
\newblock Last-level cache side-channel attacks are practical.
\newblock In {\em 2015 IEEE Symposium on Security and Privacy}, pages 605--622.
  IEEE, 2015.

\bibitem{cacti}
N.~Muralimanohar, R.~Balasubramonian, and N.~P. Jouppi.
\newblock Cacti 6.0: A tool to model large caches.
\newblock {\em HP laboratories}, 27:28, 2009.

\bibitem{dissertation_divya}
D.~Ojha.
\newblock {\em Redesigning Caches to Resist Side Channel Attacks}.
\newblock PhD thesis, University of Rochester, 2022.

\bibitem{timecache}
D.~Ojha and S.~Dwarkadas.
\newblock {TimeCache: Using Time to Eliminate Cache Side Channels when Sharing
  Software}.
\newblock In {\em 2021 ACM/IEEE 48th Annual International Symposium on Computer
  Architecture (ISCA)}, pages 375--387. IEEE, 2021.

\bibitem{primeprobe}
D.~A. Osvik, A.~Shamir, and E.~Tromer.
\newblock Cache attacks and countermeasures: the case of aes.
\newblock In {\em Cryptographers’ track at the RSA conference}, pages 1--20.
  Springer, 2006.

\bibitem{probpnp}
A.~Purnal and I.~Verbauwhede.
\newblock Advanced profiling for probabilistic prime+ probe attacks and covert
  channels in scattercache.
\newblock {\em arXiv preprint arXiv:1908.03383}, 2019.

\bibitem{ceaser}
M.~K. Qureshi.
\newblock Ceaser: Mitigating conflict-based cache attacks via encrypted-address
  and remapping.
\newblock In {\em 2018 51st Annual IEEE/ACM International Symposium on
  Microarchitecture (MICRO)}, pages 775--787. IEEE, 2018.

\bibitem{ceasers}
M.~K. Qureshi.
\newblock New attacks and defense for encrypted-address cache.
\newblock In {\em Proceedings of the 46th International Symposium on Computer
  Architecture}, pages 360--371. ACM, 2019.

\bibitem{vway}
M.~K. Qureshi, D.~Thompson, and Y.~N. Patt.
\newblock The v-way cache: demand-based associativity via global replacement.
\newblock In {\em 32nd International Symposium on Computer Architecture
  (ISCA'05)}, pages 544--555. IEEE, 2005.

\bibitem{rane15raccoon}
A.~Rane, C.~Lin, and M.~Tiwari.
\newblock Raccoon: Closing digital side-channels through obfuscated execution.
\newblock In {\em 24th $\{$USENIX$\}$ Security Symposium ($\{$USENIX$\}$
  Security 15)}, pages 431--446, 2015.

\bibitem{rane16}
A.~Rane, C.~Lin, and M.~Tiwari.
\newblock Secure, precise, and fast floating-point operations on x86
  processors.
\newblock In {\em 25th $\{$USENIX$\}$ Security Symposium ($\{$USENIX$\}$
  Security 16)}, pages 71--86, 2016.

\bibitem{mirage}
G.~Saileshwar and M.~Qureshi.
\newblock $\{$MIRAGE$\}$: Mitigating conflict-based cache attacks with a
  practical fully-associative design.
\newblock In {\em 30th $\{$USENIX$\}$ Security Symposium ($\{$USENIX$\}$
  Security 21)}, 2021.

\bibitem{fantastic}
M.~Schwarz, C.~Maurice, D.~Gruss, and S.~Mangard.
\newblock Fantastic timers and where to find them: high-resolution
  microarchitectural attacks in javascript.
\newblock In {\em International Conference on Financial Cryptography and Data
  Security}, pages 247--267. Springer, 2017.

\bibitem{bayes}
R.~Swinburne.
\newblock Bayes' theorem.
\newblock {\em Revue Philosophique de la France Et de l}, 194(2), 2004.

\bibitem{phantom}
Q.~Tan, Z.~Zeng, K.~Bu, and K.~Ren.
\newblock Phantomcache: Obfuscating cache conflicts with localized
  randomization.
\newblock In {\em NDSS}, 2020.

\bibitem{evictionset}
P.~Vila, B.~K{\"o}pf, and J.~F. Morales.
\newblock Theory and practice of finding eviction sets.
\newblock In {\em 2019 IEEE Symposium on Security and Privacy (SP)}, pages
  39--54. IEEE, 2019.

\bibitem{secdcp}
Y.~Wang, A.~Ferraiuolo, D.~Zhang, A.~C. Myers, and G.~E. Suh.
\newblock Secdcp: secure dynamic cache partitioning for efficient timing
  channel protection.
\newblock In {\em Proceedings of the 53rd Annual Design Automation Conference},
  page~74. ACM, 2016.

\bibitem{newcache}
Z.~Wang and R.~B. Lee.
\newblock New cache designs for thwarting software cache-based side channel
  attacks.
\newblock {\em ACM SIGARCH Computer Architecture News}, 35(2):494--505, 2007.

\bibitem{scattercache}
M.~Werner, T.~Unterluggauer, L.~Giner, M.~Schwarz, D.~Gruss, and S.~Mangard.
\newblock Scattercache: thwarting cache attacks via cache set randomization.
\newblock In {\em 28th $\{$USENIX$\}$ Security Symposium ($\{$USENIX$\}$
  Security 19)}, pages 675--692, 2019.

\bibitem{PaaS}
Y.~Zhang, A.~Juels, M.~K. Reiter, and T.~Ristenpart.
\newblock Cross-tenant side-channel attacks in paas clouds.
\newblock In {\em Proceedings of the 2014 ACM SIGSAC Conference on Computer and
  Communications Security}, pages 990--1003, 2014.

\end{thebibliography}

\end{document}